# A SECURITY EVALUATION FRAMEWORK FOR U.K. E-GOVERNMENT SERVICES AGILE SOFTWARE DEVELOPMENT

Steve Harrison[1], Antonis Tzounis[2], Leandros Maglaras[1], Francois Siewe[1], Richard Smith[1] and Helge Janicke[1]

[1]De Montfort University, The Gateway, Leicester LE1 9BH, United Kingdom
[2]Department of Agriculture, Crop Production & Rural Environment, University of Thessaly, Volos, Greece

*ABSTRACT*

*This study examines the traditional approach to software development within the United Kingdom Government and the accreditation process. Initially we look at the Waterfall methodology that has been used for several years. We discuss the pros and cons of Waterfall before moving onto the Agile Scrum methodology. Agile has been adopted by the majority of Government digital departments including the Government Digital Services. Agile, despite its ability to achieve high rates of productivity organized in short, flexible, iterations, has faced security professionals' disbelief when working within the U.K. Government. One of the major issues is that we develop in Agile but the accreditation process is conducted using Waterfall resulting in delays to go live dates. Taking a brief look into the accreditation process that is used within Government for I.T. systems and applications, we focus on giving the accreditor the assurance they need when developing new applications and systems. A framework has been produced by utilising the Open Web Application Security Project's (OWASP) Application Security Verification Standard (ASVS). This framework will allow security and Agile to work side by side and produce secure code.*

## 1. INTRODUCTION

This paper will be based around three concepts; firstly a literature review based on current software development methodologies, like Waterfall and Agile, as well as a brief review of the accreditation process and the Government Service Design Manual. Secondly a gap analysis on the findings from the literature review, and, thirdly recommendations for addressing these gaps. We finally present a framework in the shape of an excel spreadsheet, which is based on the Open Web Application Security Project's (OWASP) Application Security Verification Standard (ASVS).

In the United Kingdom (U.K.) Her Majesty's Government (HMG) are making more public services available on-line for its citizens.[1] An example of this is purchasing vehicle tax online now, instead of over the counter, in a Post Office. A number of departments and offices are currently undergoing huge changes in making these services available, for example the Department of Work and Pensions (DWP) are changing the way we access parts of the benefits system by allowing claimants to make benefit claims on-line using the Universal Credit (UC) system. The security within these on-line systems is of utmost importance to prevent fraud and

International Journal of Network Security & Its Applications (IJNSA) Vol.8, No.2, March 2016

error.[2][3]Security is equally important for the front-end Internet services, as well as the back-end processing systems. Therefore, security by design during the initial requirements phase is a must.

Agile is a software development methodology that has been used in the commercial world for some time, and it comes in different methods, such as Scrum, eXtreme Programming (XP), Crystal and Adaptive Software Development (ASD). I.T. and digital departments within HMG, despite their initial objections, are nowadays following the Agile approach to software/application development. Security and Agile are not a good mix, though, and security can be seen as a blocker in some cases. The formal HMG security accreditation process does not fit well with Agile as this predominately follows a Waterfall process.[4] Agile is replacing the traditional 'Waterfall' methodology for software and digital project development. What is needed is a way of embedding security into the Agile process without slowing down the rapid development nature of Agile. At the same time, we need to give the accreditor and the senior business owner the assurance they need to formally sign off the system for live use.

The focus of this work will be on developing a security framework that can be used within Agile sprints to develop secure applications and to give assurance to both the accreditor/senior business owner that any technical risks have been mitigated.

## 2. THE WATERFALL DEVELOPMENT MODEL

Our first aim was to conduct a mapping of the pros and cons of the methodologies at hand. First of all we researched the Waterfall development method. Then we proceed with the Agile and the Scrum methodology investigation, followed by an analysis on the current HMG accreditation process and the Government Service Design Manual. Finally we investigate the use of testing frameworks within Agile sprints in particular OWASP Application Security Verification Standard (ASVS).

Traditionally, software and system development within Her Majesty's Government (HMG) Departments and Offices has followed a Waterfall methodology for development of anything from small to large I.T. projects that have an impact on whole organisations for example the National Health Service (NHS). The Waterfall methodology involves a series of cascading steps that cover the development process with a small level of iteration between each stage. The major problem with using the Waterfall methodology for the development of Web applications (and also Information Systems) is the rigidity of its structure and lack of iteration between any stage, other than adjacent stages. We should look in detail at each stage of the waterfall methodology (Figure 1).

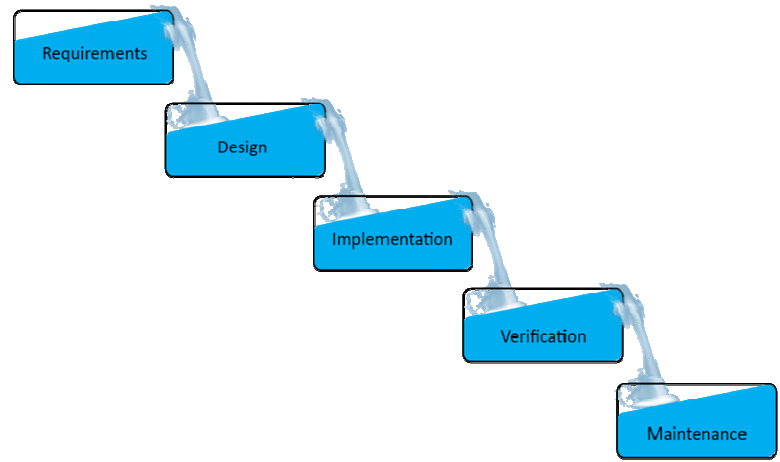

Figure 1. Waterfall methodology Model.

Waterfall methodology is organized into the next sub-processes (phases):

A) **Requirements phase:** Its main focus is to define and capture the business requirements and user needs that the system is being designed to solve. For example, take the NHS. A business requirement could be to "analyse illness trends in relation to seasons". A team of business analysts, business users, managers and I.T. experts will be created to ensure all the requirements are gathered correctly. The requirements phase, typically, consumes approximately 30% of the overall development cycle. The requirements are usually "set in stone" with little room for change once the decision has been made. Once the main requirement have been gathered then subset requirements will be generated for example the I.T. requirements could be split into security, service level agreements (SLA) or software development. [5]

B) **Design Phase:** With all the requirements gathered the design of the actual system starts to take place. Within a software project the design phase is split into two sections. The first is the system design, whichundertakes the overall system details and specifications. It also includes how each component will interact with each other by means of a data flow diagram.The second is the component design,which focuses on how each individual component will operate separately. Software engineers are assigned to components to plan how they will interact with each other. This has to be documented as this will form the input for the next phase. The design phase will use approximately 35% of the overall development cycle.[6]

C) **Implementation Phase:** This phase is when the software development actually starts. The information from the first two phases is gathered and converted into working software by means of the components.

D) **Verification Phase:** This phase is where the testing takes place. Usually the first testing is User Acceptance Testing (UAT). This is to ensure that the system actually does what the requirements and the design phases stipulated. Within HMG this is the phase that Penetration Testing or an I.T. Health Check (I.T.H.C.) would be carried out. The testing would be looking for vulnerabilities within the system infrastructure and applications

53

running the system. If major issues are found then this could result in components being re-written in software, this can have the knock on affect that if one component is changed it can affect the operation of other components. The implementation and verification phase will use approximately 30% of the overall development cycle.

E) **Maintenance Phase:** This phase of the project is normally where the system is signed off for use by an accreditor and operationally accepted by the business. This is to ensure everything is running within parameters and that changes or patches to the system are applied using change control methods. The maintenance phase is also preparing everything for "go-live", for example the training of staff, ensuring documentation is complete and handing over the system to the operational staff. The Ops staff will be responsible for the daily running of the system. The maintenance phase will use approximately 5% of the overall development cycle.

As we can see the output of each stage is the input for the following stage. However, the Waterfall method does not fit into modern software development needs were business requirements are often rapidly changing. Waterfall is often referred to a "Big Design Upfront" (BDUF) approach, this is where the application design is to be completed and perfected before the application implementation is started. Hence the need for a more flexible approach to software development and Agile [7].

The Internet is a rapidly advancing environment with new technologies becoming available almost daily. Any methodology used for the development of Web sites must be flexible enough to cope with change.[8]. This does not only apply to web sites but can also apply to I.T. projects in general.In 1970 Winston W. Royce delivered a paper to the IEEE WestCom engineering conference, this paper described what is now considered traditional waterfall. The paper described a sequential process where each phase is completed before the next begins. Royce offered this model as an example of how not to do software development. However, the audience liked this development model and in 1985 the American Department of Defence (DoD) adopted it as the official methodology for developing projects. [9]

**2.1. Waterfall Issues**

Waterfall follows a sequential process completing each phase before moving onto the next. Each phase is fully documented before moving on, this documentation can take a long time to achieve resulting in projects being delayed. The production of vast amounts of documentation during the initial requirements phase can lead to the omissions (due to the process being tedious) of some requirements which can have a serious knock on effect during the other phases of the process.
During Waterfall development the requirements both business and system are "set in stone". Indeed, the backbone to Waterfall is in the requirement phase. This freezing of the requirements is great for software developers as everybody knows upfront what is expected. However, technologies are changing on a daily basis. Moreover, business or customers may not know the exact requirements they have. So, software development has to be dynamic and adapt to these changes and system requirements. Waterfall does not accommodate these changes readily.

By the time the system reaches operational 'go live' and is about to be handed over to the operations team, the systems are in desperate need for software and patching updates. This again can add to the delay for the project to go live due to operations rejecting a system that is needing patching. There is normally a discussion or quite often a standoff between operations and project staff about the handover, with the operations staff having to take on the operational running of the system due to pressure from the business. Normally at the end of a Waterfall project there is a "lessons learnt" meeting, which can, and often, lasts for hours. It is very much focused on team/technology/project specific issues that have occurred during the project. In the authors experience the majority of people who attend a "lessons learnt" meeting that lasts for several hours are reluctant to mention issues they encountered for the sheer fact they want to get out of the meeting and move onto the next project.When Government has used Waterfall in the past this has often resulted in huge overspend, delays in getting the system operational and less functionality than originally expected.

## 2.2. Agile Scrum

Scrum is a simple framework used to organize teams and get work done more productively, with higher quality. It is a "lean" approach to software development that allows teams to choose the amount of work to be done and decide how to do it best. Designed to adapt to changing requirements during the development process at short, regular intervals, it prioritizes customer requirements to evolve the work product in real time to customer needs. In this way, Scrum provides what the customer wants at the time of delivery (improving customer satisfaction) while eliminating waste (work that is not highly valued by the customer) [10]. Although Sutherland gives a very good description of Scrum there is no mention of security considerations when developing software.

Agile follows a thought process of Fail Fast, Fail Often in order to improve the software and the teams developing the software. Many agile puritans read these agile values and interpret them incorrectly, the use of the word "over" is misconstrued to mean "instead of" an example this would be the second agile value "Working software over comprehensive documentation", or, "Working software" is more valuable than "comprehensive documentation" when thinking about delivery to the customer in principle number one.

### 2.2.1. Scrum

It has been said that Agile Scrum is compared to a rugby team scrum in where the team move and work as one. When this is compared to the relatively linear approach of Waterfall it is easy to see how Agile Scrum became known as Scrum[10][11].

The artefacts listed below are what enable the Scrum process to deliver products:

A) **Product Backlog:** A list of deliverables for the project, like: features, functionality or bug fixes.

B) **Sprint Backlog:** A list of tasks or user stories that have been identified by the Scrum team that will be completed by the sprint team.

C) **Burn Charts:** There are two type of burn charts "burn up" and "burn down". Burn charts show the team the relationship between time and scope. Time is on the horizontal X-axis while scope is on the Y-axis. A burn up chart shows how much of the scope the team has completed over a period of time. A burn down chart show what work is left to do.The two charts are used independently.

- D) **Task Board:** The task board in its simplest form consists of three columns these are: To Do, Doing, and Done.

- E) **User Stories:** Agile user stories are a main element in the methodology, they are short descriptions from the viewpoint of the user. They normally consist of the following short sentences, like: "As a <type of user>, I want <some goal> so that <some reason>"

## 2.2.2. Roles within Scrum

Scrum recognises three primary roles:

- A) **Product Owner:** Usually a member of the business who understands what is trying to be achieved by completing the project, they often have direct contact with the customers who will use the software/system.
- B) **Scrum Master:** Brings leadership to the team, but this not is leadership though influence of being a higher rank within the organisation but rather has a helpful friend or "agony aunt".
- C) **Team Member:** The scrum team usually consists of approximately 7 members, plus or minus two. The members have a variety of skills depending on the project they are taking part in.

## 2.3. Sprints

Sprints are one of the fundamental concepts of Agile, this is the process of splitting or dividing your overall project into smaller pieces. An example of this can be an application that has various functions take Microsoft Word for instance. The overall project would be to design a word processing application however functions like save, print and labels would be split into smaller pieces of work or sprints. These functions when combined with other functions are what make up the application. The same would happen in an Agile project.

### 2.3.1. Sprint Planning

Sprint planning occurs at the beginning of the sprint; the meeting is normally split into two parts. The first part of the meeting deals with the deliverables for the sprint so that the team are committed to producing what is needed. At the second part of the meeting the team deals with the identification of the tasks that are needed to complete the stories. The user story could be "as a user I want to be able to print from the application". The tasks associated with the user story could be: design the icon and design the print screen wireframe once the icon is pressed.

### 2.3.2. Sprint Review

At the end of the sprint a review takes place this review goes under the guises of "show and tell" or "sprint demo" in essence it is the same thing. It is a chance for the team members to show off their work to the stakeholders and also report on which work did not get completed.

### 2.3.3. Retrospective

This is a review meeting also held at the end of each sprint, it is a kind of lessons learn exercise that all sprints normally hold. The difference in Agile is the lessons learnt will be applied to the next sprint whereas with a Waterfall project the lessons learnt is at the end of the project with the intention of applying these lessons to the next big project. In reality this seldom happens and everybody is just happy to get out of the meeting to start another project.

**2.4. Daily scrum**

The daily scrum or stand-up meeting is as the name suggests daily meetings for team members, scrum master and product owners[12]. These are often referred to as "committed" members. Other members are "involved" members these could be sales directors, Chief Technology Officer (CTO) or Chief Information Security Officers (CISO). The meeting are normally held in the same place at the same time, each team member will be asked to inform the meeting members of the following:

A) What I accomplished since the last scrum meeting.
B) What I expect to accomplish by the next scrum meeting.
C) What obstacles are slowing me down.

Should there be any slippage or delays then the product owner will know immediately and may be able to take corrective action. In theory the product owner should be aware of all outstanding issues no more than a day old.

As we can see Agile is clearly about producing good code in a timely manner of which it does without doubt. One issue that is on most security consultants and accreditors minds is "where is the security in all of this rapid development and how do I get it accredited".

**2.5. Issues with Agile Scrum**

When comparing Agile scrum to the Waterfall methodology there are obvious advantages in favour of Agile. Being able to change user requirements for example, however all is not plain sailing within the Agile methodology.

Looking at the values of agile we can see the four main values:

A) Individuals and interactions over processes and tools
B) Working software over comprehensive documentation
C) Customer collaboration over contract negotiation
D) Responding to change over following a plan

Many agile puritans can take the above values literally and in the case of value number 2 can use this as an excuse not to document their work. Some developers see the documentation of their work as adding comments to the code to help explain certain functionality. In a study by Juyun Cho in 2008 were he interviewed nine developers in a company that produced small to medium web based projects, he found that several developers would comment their code but several did not.[13] This can lead to issues when new members join the team and are trying to understand what has been done in the past. This with the lack of a security framework can make developers write code that is unstable and insecure.

There is one gaping hole in the Agile methodology and that is lack of any security consideration.[14]The growing trend towards the use of agile techniques for building web applications means that it is essential that security engineering methods are integrated with agile processes.[15] Agile rapidly produces code on daily/weekly basis and that is what it is excellent at. However, we then revert back to the Waterfall methodology when it comes to performing an I.T.H.C. and technical accreditation activities.The Fail Fast, Fail Often is good when developing software, however this cannot be applied in a security context. This would result in data security breaches, loss of information and have massive reputational damage to the organisation who implemented it.

As we can see the Agile methodology for developing software by creating smaller chunks of work rather that an "all or bust" mentality found in Waterfall has obvious advantages. The lack of security within Agile is not entirely true, as we do have pair programing giving more assurance than security.  As we have seen this can be costly to an organisation having to double up on staff and in times were public services are being cut and civil servants are losing their jobs does not bawd well.

Performing pen testing activities within the sprint could prove to be extremely expensive when using third party testers. There is also the issue of getting pen testers on-site as there is usually a high demand for their skills within the market. Having pen testers assigned to projects to perform testing in the sprints for maybe 2 hours per day is also a waste of resource. Another solution is needed.

**2.6. Government Service Design Manual**

The following section briefly explores the Government Service Deign Manual, which is the standard that all new digital Government services are designed around. It is important to have a brief overview of this standard as this affects the accreditation process.[16]

Software development goes through these phases:

A) **Discovery:** This phase can be seen as a scoping phase where the project team is looking at user and business requirements, policies that can affect the service. If this is the transformation of an existing service into a digital service, then understanding the old service for example legacy interfaces and its underlying infrastructure is essential.

B) **Alpha:** This is the phase were the Agile SDLC is started. The alpha release is the first release of the software or prototype the idea being that it will be used by stakeholders or end users to do the following[17]:

   a. Gain an insight into the service being developed.
   b. Testing the design concepts and the technology.
   c. Building a team.
   d. Gain an understanding who or what you'll need for the Beta stage.

C) **Beta:** The objective of the beta phase is to build a fully working prototype which you can test using your end users. Within the beta phase you are continuously tweaking the by rewriting code or replacing code to ensure the prototype is ready to go live. During beta

58

> you will also ensure that any interfaces with other systems/application are operating correctly. The Beta is also were the security accreditation work is started.
>
> D) **Live:** At this stage the application is ready for release it has undergone testing and been signed off by the business that they accept any risks. Also at this stage the project team would hand over to operational support including security operations who will monitor the application and be responsible for its day to day running and keeping it secure for example via patching. It is at this phase were you get the system accredited meaning the business has accepted any residual risks and has signed this off.
>
> E) **Retirement:** At this stage the system has served its life span and will be decommissioned. Users will be informed that the service is ending. URL's will be redirected to the new service if applicable.

From a security viewpoint, the following will apply but there could be others depending on the service:

- A) Data retention, how long do we need to keep the old data.
- B) Transferring the old data to a new provider in a secure manner.
- C) Data destruction, what is a suitable method to destroy the data.
- D) Decommission of old equipment in particular storage devices

### 2.7. Issues with the Government Digital Service Manual

The Government Digital Service Manual is a relatively new process and is being fine-tuned constantly. There is potential for improvement from a security perspective.

Security is not officially engaged until the Beta phase of the service design process. Then we expect the security architect/consultant to get up to speed with the project. The security architect has then to understand the business requirements along with the application being designed and coded. The architect will start asking questions for example:

- A) What is the risk appetite of the organisation?
- B) Who are the threat actors?

A security architect/consultant will not only be looking at technical security of the service but also the legal aspects of the service for example the U.K. Data Protection Act (DPA) and other European Union (EU) legislation.

As stated above this is a new process and as the process evolves it is being improved with each iteration.

### 2.8. HMG Security Standards and the Accreditation Process

HMG is governed by a series of security standards and frameworks from a multitude of sources including the Cabinet Office (CO) and the Government Communications Headquarters (GCHQ) Information Assurance (IA) branch, Communications Electronic Security Group (CESG). CESG provide IA assistance to Government via its internal staff, publications and until recently a body of approximately 600 private sector security consultants who make up the CESG Listed Advisor Scheme (CLAS). CLAS consultants typically advise Government organisations on behalf of

International Journal of Network Security & Its Applications (IJNSA) Vol.8, No.2, March 2016

CESG on matters of IA and Security in general. The CLAS scheme is due to close between the end of 2015 and mid-2016. A new scheme called the Cyber Security Consultancy will replace it. Some examples of the many policies and frameworks that provide IA governance within HMG are:

**A)** The Security Policy Framework (SPF 2014).
**B)** CESG policies and guidelines for IA and risk management.
**C)** Data Protection Act (DPA).
**D)** Official Secrets Act (OSA).
**E)** CPNI Advice.
**F)** Council of the European Union (EU) Security Committee.

Other non U.K. Government agencies that provide advise are:
**A)** European Union Agency for Network and Information Security (ENISA).
**B)** National Institute of Standards and Technology (NIST) based in North America

Before discussing the accreditation process, it is important to discuss the accreditor. This is the person who will make decisions on behalf of the business risk owner, for example the risk owner could be a senior civil servant who is the sponsor for the project or the information asset owner. The accreditor must have a very good understanding of the business objectives; the value of the data the organisation is trying to protect.

Below is definition of accreditation this is taken from the document produced by CESG titled "CESG IA Top Tips 2014/01 Accreditation":

*"Accreditation is a decision - made by the business - to demonstrate confidence that the risks of engaging in an activity are balanced against the expected benefits of that activity."*[18]

For example, a potential supplier of a web application whose users accessed via the Internet, the application had known security vulnerabilities that could not be mitigated. The accreditor normally would not recommend the application to be used in a live production service. In a different scenario, that very same application, does not connect to the Internet, and only one person can access the application from a closed network, that has no external links to other networks and is totally isolated. In this case, the accreditor may allow the system to be used as the risk of attack via an external attack vector is far less than that of the Internet connected application.

Risk-based decisions should also take the financial costs to secure a system into consideration, if the costs to secure a system outweigh the costs that the system will generate then it is clearly not acceptable to spend the money securing the system an alternative should be sought.[19]The accreditation process is often lengthy and heavily dependent on documentation being produced at every stage and generally follows the Waterfall methodology. Accreditation is often engaged well after the system has been designed and is ready to go into production (live), this can produce extra costs to the business in having to redesign elements of the system to gain accreditation. It is often heavily dependent on the production of Risk Management and Accreditation Documentation Sets (RMADS). This is a document the accreditor will sign off to show he/she is happy with the risk approach that has been taken to the system. RMADS can be 150 plus pages' long that are stored in a secure repository gathering dust and will generally only come out once a year for review and an I.T.H.C. (IT Health Check) being carried out. In between that time, we are reliant on a good patch management process keeping systems updated from known

International Journal of Network Security & Its Applications (IJNSA) Vol.8, No.2, March 2016International Journal of Network Security & Its Applications (IJNSA) Vol.8, No.2, March 2016 vulnerabilities. In essence the RMADS are a snap shot of the system at a singular point in time, in today's fast paced world a more dynamic process is needed to keep pace with technology and attack vector changes.

Normally within the accreditation process a penetration test or I.T.H.C. is carried out on the post live application. For HMG this is normally carried out by a third party testing company who a CHECK accredited by CESG allowing them to penetration test HMG systems and applications.Once the I.T.H.C. has been carried out a report is generated on the vulnerabilities found within the application. This will form a basis for the production of a risk treatment plan (RPT), the RTP will list in order of severity the vulnerabilities found. The organisation will have a risk appetite; this is the amount of risk the organisation is prepared to tolerate before allowing a solution to go live. This is normally either set at High, Medium or Low. Once a vulnerability has been treated and mitigated the next vulnerability will be treated, this process goes on until all the risk at the risk appetite and above have been mitigated. Once this has happened and the risk management documentation has been complied and accepted will the system be allowed to go live. There are exceptions to this within some Government departments. The accreditor and the business can grant an Authority to Operate (ATO) allowing the system to be tested prior to the formal accreditation process to be completed.

The accreditation process is heavily reliant on documentation, while the Agile approach supports the idea of only producing documentation when necessary and being ready to incorporate changes rapidly. Obviously, the two approaches do not bond very well.

**2.9. Issues with the Accreditation Process**

The accreditation process currently used with HMG is not suitable for a fast-paced methodology such as Agile. This is due to the accreditation process generally following a Waterfall methodology and not being easily adaptable to rapid changes.As we have seen, Agile is about creating software and getting it live as soon as possible. Within the Agile principles, there is no mention on how we create secure software; this is left to the individual organisation to try and resolve.

The accreditation process differs between different projects. There is not a standard or a "one size fits all" approach. Each project has to be assessed within its own rights.What is needed is a new accreditation methodology that can work well with Agile but still satisfy the requirements of the business and accreditor to show that risks and in particular technical risks have been mitigated to an acceptable level for the business to be satisfied.

It could be that we are now entering a new way of trying to ensure our systems and software are secure, by using assurance rather than a formal accreditation process.

## 3. OWASP APPLICATION SECURITY VERIFICATION STANDARD

One of the aims of the OWASP Application Security Verification Standard (ASVS) project is to enable a framework for performing Web application security verification using a commercially workable open standard that anybody can contribute to. The standard provides a basis for testing web application technical security controls, as well as any technical security controls in the environment, that are relied on, to protect against vulnerabilities such as Cross-Site Scripting (XSS) and SQL injection. This standard can be used to establish a level of confidence in the

61

security of web applications which can greatly assist the accreditor in his assessment of the risks associated with the application.

ASVS can be used to produce an internal or external I.T.H.C. testing scope document. Third party and internal testers can test the application using the scope document ASVS produces. The standard will also provide guidance to application developers as to what security considerations to think about when developing the code for the application. This is also known as "security by design". Incorporating "security by design" can save time and work having to retrofit and fix issues that are otherwise highlighted in the I.T.H.C. or pen test. It also gives an accreditor confidence that applications/systems are being developed with due diligence. ASVS can also assist in the writing of contracts or tender documents, to ensure suppliers are aware of what security controls are needed within the application they are developing. (OWASP, 2014)

ASVS uses three levels for security controls, these are Levels 1, 2 and 3. The definition of these levels is as follows:

**A)** L1 is intended for all software.
**B)** L2 is for applications that process sensitive data that requires protection.
**C)** L3 is for systems that handle sensitive personal data and or data that could have an impact on national security.

The table below is shows the industry and threat profile it also gives examples of the three levels discussed above. This has the potential to be mapped to HMG security classifications:

**A)** L1 (OFFICIAL)
**B)** L2 (OFFICIAL/OFFICIAL SENSITIVE)
**C)** L3 (SECRET/TOP SECRET)

Table 1. ASVS Levels. [20]

| Industry | Threat Profile | Recommendation | | |
|---|---|---|---|---|
| | | L1 | L2 | L3 |
| **Manufacturing, professional, transportation, technology, utilities, infrastructure, Government and defence.** | These industries may not appear to have very much in common, but the threat actors who are likely to attack organizations in this segment are more likely to perform focused attacks with more time, skill, and resources. Often the sensitive information or systems are not easy to locate and require leveraging insiders and social engineering techniques. Attacks may involve insiders, outsiders, or be collusion between the two. Their goals may include gaining access to intellectual property for strategic or technological advantage. We also do not want to overlook attackers looking to abuse application functionality influence the behavior of or disrupt sensitive systems. Most attackers are looking for sensitive data that can be used to directly or indirectly profit from to include personally identifiable information and payment data. Often the data can be used for identity theft, fraudulent payments, or a variety of fraud schemes. | All network accessible applications. | Applications containing internal information or information about employees that may be leveraged in social engineering. Applications containing nonessential, but important intellectual property or trade secrets. | Applications containing valuable intellectual property, trade secrets, or government secrets (e.g. in the United States this may be anything classified at Secret or above) that is critical to the survival or success of the organization. Applications controlling sensitive functionality (e.g. transit, manufacturing equipment, control systems) or that have the possibility of threatening safety of life. |

The thought process behind the spreadsheet is for it to act as a framework providing guidance for security architects and developers when developing applications within an Agile sprint. The framework can also provide guidance for creating testing scope documents and for commercial departments to provide a baseline of security controls when outsourcing application development.

### 3.1. How to Use the Framework Spreadsheet

The screenshot below (Figure 2) shows the first worksheet that you will come to on opening the workbook. The "Cover" worksheet is the default, you must Enable Content that is displayed in the Security Warning to use the workbook. Here you can see the list of security controls embedded within the buttons.

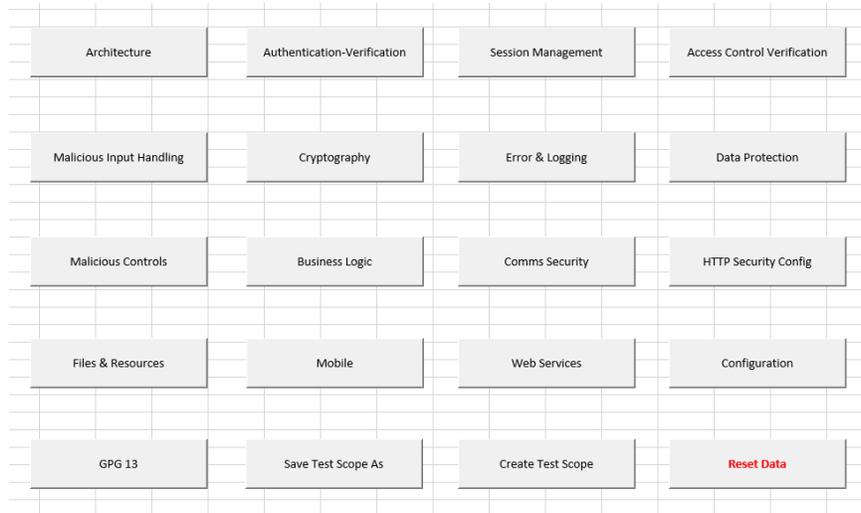

Figure 1. Initial Screen.

The first action is press the "Reset Data" button this will zero all entries on all worksheets ready for data input.

Figure 2. Security Control Screen.

The next thing to do is select the security controls you require for your project. You can use "Y", "y", "YES", "yes" or "Yes". Note you need to select the section, this example "Architecture Design and Threat Modelling" has a Y in the "Required" cell. After you have selected the controls you need click the "Back to Cover Sheet" button and select the next security controls needed. Do this until all required controls have been selected. With the required security controls selected next select the "Create Test Scope" button this will then create the required test scope as in figure 4.

Figure 4. Test Scope.

International Journal of Network Security & Its Applications (IJNSA) Vol.8, No.2, March 2016Once the test scope is created you have the option to save the test scope worksheet into a separate workbook. You can use the "Save Test Scope As" button on the "Test Scope" worksheet or go back to the "Cover" worksheet to save. You can change the name of the file to be saved to your own choice. The test scope is created in the same directory as the framework spreadsheet.To create another test scope you will need to reset the data and start the process again.

## 4. RECOMMENDATIONS

There are three main common uses for the framework are:
- **A)** To assist security architects in writing I.T.H.C/Pen testing scope documents.
- **B)** To assist developers within Agile sprints.
- **C)** To assist Commercial departments in the tender process.

What will be discussed below is a more detailed look into each of the three uses.

### 4.1. Security by Design within Agile Sprints

By incorporating the security framework within the Agile SDLC the security architect along with the software developers can help bridge the gap that exists within Agile Scrum in the fact that there are no security considerations.The framework could be discussed within the sprint backlog meeting with the security architect present. In a joint effort the security architect and developers can go through the framework and match the controls to the sprint. This has the added benefit that the security architect is now a sprint team member and included within the team. What has happened in the past is that the security architect is often seen as a non-team member and communication between the developers and the architect is non-existent. This also fits with the following Agile values and principles.

**Values:**

- **A)** Individuals and interactions over processes and tools.

**Principles:**

- **A)** Business people and developers must work together daily throughout the project.
- **B)** The most efficient and effective method of conveying information to and within a development team is face-to-face conversation.
- **C)** Continuous attention to technical excellence and good design enhances agility.

In addition to using the framework within the Agile sprints, automated source code checking could be used. This would give some form of assurance to the business that the application has no glaring security holes in it. It would also aid in situations where the infrastructure has not been built to test the code on.A copy of the test scope should also be placed next to the Agile task board acting as a daily reminder of the security controls for the sprint.

### 4.2.I.T.H.C/Pen Testing Development

The framework in this instance could be used to help the security architects in developing an internal testing scope or a testing scope for third party testers.Security architects and consultants need to have an understanding of the ongoing work, and be able to perform penetration testing to

65

a suitable standard. The framework should be used as the scope for such testing. Having a security architect available to security test code would enable the smooth integration of security within the Agile sprint.

The application will still have to be formally penetration tested by a CHECK test team prior to going live (only in Central Government) but by testing internally we can be assured that no huge security vulnerabilities exist. This will reduce the risk that the final testing will find any vulnerabilities that could prevent the application from go live.

### 4.3. Commercial Tender Process

When outsourcing the development of systems to third parties the framework can be used as a baseline of security controls that must be followed by the supplier's developers. Suppliers can add addition security controls but must adhere to the framework as a reference baseline.

### 4.4. Government Service Design Manual

In the Government Service Design Manual security and accreditation are not engaged until the Beta phase of the process, what would be better is rather than having one work stream have two. The first work stream will follow the process as advised in the manual. The second work stream would be the security work stream. This would run in parallel when the discovery phase begins. This would enable security to have a greater understanding of the business requirements and current issues. It would also enable security to start on the risk and threat modelling activities, these would be fine-tuned and updated until the Beta phase begins.

### 4.5. General Recommendations

The framework could be used as a reference when organisations are developing applications using non Agile methodologies, for example the framework could be used as a reference artefact for use within The Open Group Architecture Framework (TOGAF).

Security architects and security consultants often come from a networking or infrastructure background, with smaller software development experience. It would be impossible for a security architect to be proficient in every software development language for the purpose of security checking code. What could be achieved is providing basic security training or briefings to developers to give them an understanding of what security concerns they should be addressing within the sprint. The developers are at the coal face and working with code on a daily basis, whereas a security architect or consultant may not be.In addition to using the framework within the Agile sprints, automated source code checking could be used. This would give some form of assurance to the business that the application has no glaring security holes in it.

## 5. CONCLUSION

This work takes an initial step into the integration of a security framework into Agile scrum used within HMG. For Agile and the security accreditation processes to work together there has to be a compromise between the two. For a long time, security has been seen as a blocker or at least a massive speed bump that slows down a project, in some cases bringing the project to a complete halt.

Project managers and the business will often try to circumvent the speed bump, often by deliberately not informing the security team about decisions made within the project. This can, and often does, have the undesired effect of weakening the security and allowing vulnerabilities to be missed.For systems at the OFFICIAL classification, the "old style" way we think about security within HMG has to change. There still has to be a formal process of providing evidence to the business that a system or application is as secure as it can be. However, this process is moving more to an assurance rather than full blown accreditation process.

The business needs to engage with security at the very beginning of the discovery phase and not at the very end of the development process as often this is too late for security to have any impact. Security has to be seen as a business enabler and be embedded into the overhaul process, not something that we think about at the last minute.

**Authors**

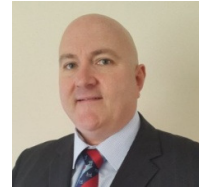

Mr. Steve Harrison is a Senior Enterprise Security Architect, specializing in I.T. security infrastructure and Information Assurance. He is employed by CyberSecurity Consultants Ltd based in the United Kingdom, often employed on Government I.T projects as a security advisor. Steve is a Certified CESG Professional (CCP) at senior IA architect level, a ISC2 Certified Information System Security Professional (CISSP) and a ISC2 certified Information Systems Security Architecture Professional (ISSAP).

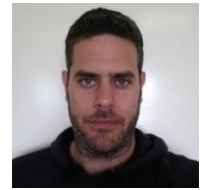

Antonis Tzounis received his B.Sc. degree in Cultural Technology and Communication (Cultural Informatics) of the University of the Aegean and MSc from the department of Electrical Engineering, University of Thessaly, Greece. As a PhD student of the Agricultural School of the University of Thessaly, he is currently involved in several research projects emphasizing on the design and development of IoT/WSN deployments and web services for surveillance, monitoring and control applications, focusing on agricultural sector research. His research interests include embedded systems programming and communications, sensors, greenhouse climate control and modelling.

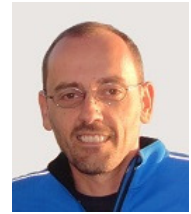

Dr. LeandrosMaglaras received the B.Sc. degree from Aristotle University of Thessaloniki, Greece in 1998, M.Sc. in Industrial Production and Management from University of Thessaly in 2004 and M.Sc. and PhD degrees in Electrical & Computer Engineering from University of Volos, in 2008 and 2014 respectively. He is currently a Lecturer in the School of Computer Science and Informatics at the De Montfort University, U.K. During 2014 he was a Research Fellow in the Department of Computer Science at the University of Surrey, U.K. He has participated in various research programs investigating vehicular and ICT technologies (reduction-project.eu), sustainable development (islepact.eu), cyber security (cockpitci.eu, fastpass-project.eu) and optimization and prediction of the dielectric behavior of air gaps (optithesi.webs.com). His research interests include wireless sensor networks and vehicular ad hoc networks. He is an author of more than 40 papers in scientific magazines and conferences.

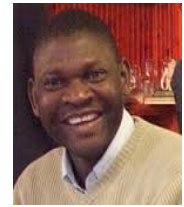

Dr Francois Siewe is a Senior Lecturer in the Software Technology Research Laboratory (STRL) of the Faculty of Technology at De Montfort University (DMU), in Leicester in the UK. Before joining STRL, he was a research fellow on the EPSRC-funded project MELANGE on modelling the structure dependent colour properties of melange yarns in the Textile Engineering And Material (TEAM) research group at DMU. This followed tenure as lecturer and vistiting researcher in the Institute of Technology of Lens at University of Artois in Lens, France. Prior to this, he was a fellow at the United Nation University/ International Institute for Software Technology (UNU/IIST) in Macau, where he worked on the Design Techniques for Reat-time systems (DeTfoRs) project. He was also a lecturer in the Department of Mathematics and Computer Science at the University of Dschang, Dschang, Cameroon.

Dr Richard Smith is a Senior Lecturer at De Montfort University. He is a member of the CSC/STRL and has worked extensively with the European Space agency and NASA, as both technical lead and prime for global international consortia. He has managed contracts worth over €2 million with teams comprising partners from both academia and industry, producing scientific results far exceeding original expectations which led to numerous CCNs to expand the remit of projects such as River & Lake, hosted by DMU on behalf of ESA. Richard has published over 40 papers in peer-reviewed publications and has presented at numerous international conferences to world experts.

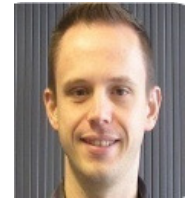

Dr. Helge Janicke obtained his first degree in "practical informatics" from the University of Applied Sciences, Emden (Germany). During his doctoral studies he was funded by the Data and Information Fusion Defence Technology Centre (DIF-DTC), a research consortium of high-tech companies and universities which formed a key plank of the UK Government's future vision for defence technology development. He was awarded his PhD in 2007 from De Montfort University (DMU) and subsequently worked for the DIF-DTC consortium as a Research Fellow, funded jointly by QinetiQ and the Ministry of Defence. In 2008, Janicke worked for the University of Leicester as a Teaching Fellow leading several modules on software engineering, quality assurance and measurement theory. He provided consultancy services to SGS/Ofgem on quality assurance and testing in software used in the UK's gas supply network. Janicke worked on the NATO funded project "Trust Management in Networks of Networks" in collaboration with the University of Maryland (US)and the University of Skopje (FYROM). In January 2009, Janicke returned to DMU to lead the Computer Security and Trust research theme in the Software Technology Research Laboratory (STRL).

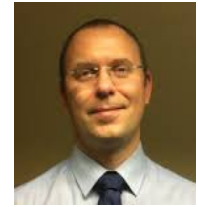